# Photovoltaic and Photothermoelectric Effect in a Double-Gated WSe$_2$ Device

*Dirk J. Groenendijk*[*,†], *Michele Buscema*[†], *Gary A. Steele*[†], *Steffen Michaelis de Vasconcellos*[‡], *Rudolf Bratschitsch*[‡], *Herre S.J. van der Zant*[†] *and Andres Castellanos-Gomez*[*,†,+]

[†]   Kavli Institute of Nanoscience, Delft University of Technology, Lorentzweg 1, 2628 CJ Delft (The Netherlands).
[‡]   Institute of Physics, University of Münster, D-48149 Münster (Germany).
[+]   Present address: Instituto Madrileño de Estudios Avanzados en Nanociencia (IMDEA-Nanociencia), 28049 Madrid (Spain)

*Dirk J. Groenendijk d.j.groenendijk@student.tudelft.nl
*Andres Castellanos-Gomez a.castellanosgomez@tudelft.nl , andres.castellanos@imdea.org

ABSTRACT

Tungsten diselenide (WSe$_2$), a semiconducting transition metal dichalcogenide (TMDC), shows great potential as active material in optoelectronic devices due to its ambipolarity and direct bandgap in its single-layer form. Recently, different groups have exploited the ambipolarity of WSe$_2$ to realize electrically tunable PN junctions, demonstrating its potential for digital electronics and solar cell applications. In this Letter, we focus on the different photocurrent generation mechanisms in a double-gated $WSe_2$ device by measuring the photocurrent (and photovoltage) as the local gate voltages are varied independently in combination with above- and below-bandgap illumination. This enables us to distinguish between two main photocurrent generation mechanisms: the photovoltaic and photothermoelectric effect. We find that the dominant mechanism depends on the defined gate configuration. In the PN and NP configurations, photocurrent is mainly generated by the photovoltaic effect and the device displays a maximum responsivity of 0.70 mA/W at 532 nm illumination and rise and fall times close to 10 ms. Photocurrent generated by the photothermoelectric effect emerges in the PP configuration and is a factor of two larger than the current generated by the photovoltaic effect (in PN and NP configurations). This demonstrates that the photothermoelectric effect can play a significant role in devices based on $WSe_2$ where a region of strong optical absorption, caused by e.g. an asymmetry in flake thickness or optical absorption of the electrodes, generates a sizeable thermal gradient upon illumination.

MAIN TEXT

Two-dimensional (2D) semiconductors have recently emerged as promising candidates for next-generation electronic devices due to their strength[1], flexibility[1,2] and transparency[3,4]. These materials are particularly interesting for optoelectronics due to their direct-bandgap in single-layer form.[3] As such, 2D semiconductors could form the basis for flexible and transparent optoelectronic devices such as solar cells, LEDs, and photodetectors. However, most 2D semiconductors such as molybdenum disulfide ($MoS_2$) exhibit predominantly unipolar behavior[5-8] while an ambipolar material is required to create a PN junction, the main element of many optoelectronic devices. Tungsten diselenide ($WSe_2$), a semiconducting transition metal dichalcogenide (TMDC), has shown ambipolar performance when used as channel material for field-effect transistors (FETs).[9] Recently, electrostatically defined PN junctions in monolayer $WSe_2$ have been fabricated, demonstrating their potential for optoelectronic devices.[10-12] In these devices, a pair of local gates is used to control the carrier type and density in the $WSe_2$ flake. Applying opposite local gate voltages results in the formation of a PN junction in the channel, generating a built-in electrical potential. Under illumination, the resulting electric field can separate photoexcited carriers. Therefore, the photocurrent generation in these devices is expected to be mainly photovoltaic. On the other hand, 2D semiconductors are also expected to show a large thermopower[13-15] which may yield a zero-bias photocurrent, even under below-bandgap illumination, due to the photothermoelectric effect as observed in $MoS_2$.[16] Photocurrent generated by the photothermoelectric effect can arise when e.g. an asymmetry in flake thickness or optical absorption of the electrodes gives rise to a sizeable thermal gradient. The emergence of the photothermoelectric effect, its coexistence with the photovoltaic effect and their relative magnitudes have not yet been studied in double-gated $WSe_2$ devices.

In this Letter, we fabricate a double-gated $WSe_2$ device with a hexagonal boron nitride (h-BN) gate dielectric and characterize the device in dark and under illumination. The gate-defined PN junction displays a cut-off wavelength of 770 ± 35 nm and measure rise and fall times close to 10 ms. Measurements of the photocurrent (and photovoltage) under above- and below-bandgap illumination show that the photocurrent generation is dominated by either the photovoltaic or the photothermoelectric effect, depending on the defined gate configuration. When the device is operated in the PN or NP configuration, the photocurrent is mostly generated by the photovoltaic effect, in agreement with previously reported results on electrostatically defined $WSe_2$ PN junctions. When the gates are biased in the PP configuration, we find that photocurrent is mainly generated by the photothermoelectric effect. Interestingly, in our device geometry the photothermoelectric current (in PP configuration) is more than twice as large as the current generated by the photovoltaic effect (in PN/NP configuration).

Device fabrication starts by patterning a pair of local gates (Ti/AuPd, 5 nm/25 nm). Next, we transfer a thin (10 nm) hexagonal boron nitride (h-BN) flake on top of the local gates to function as gate dielectric, exploiting its atomically flat surface, charge traps-free interface and large dielectric breakdown electric field.[12,17] Subsequently, we transfer a single-layer $WSe_2$ flake onto h-BN. Both transfer steps are performed via a recently developed deterministic dry-transfer method.[18] Contact to the flake is made by patterning four leads (Cr/Au, 0.3 nm/60 nm). Figure 1a shows a schematic cross section of the device, indicating the different components. The two inner leads (labeled as 2,3) define a channel which consists only of single-layer $WSe_2$, whereas the outer leads (labeled as 1,4) define a channel consisting of multilayered and single-layer $WSe_2$. In the following electrical measurements, the current flow is measured in two-terminal configuration (between leads 4 and 1, grounded) unless otherwise specified. Figure 1b shows an optical micrograph of the final device where the materials, leads and gates are labeled for clarity. The fabrication steps and characterization of the exfoliated flakes are shown in more detail in Figure S1 and Figure S2 in the Supporting Information. Photoluminescence and Raman spectroscopy measurements confirm the single-layer thickness of the WSe$_2$ flake (Figure S3 in the Supporting Information).

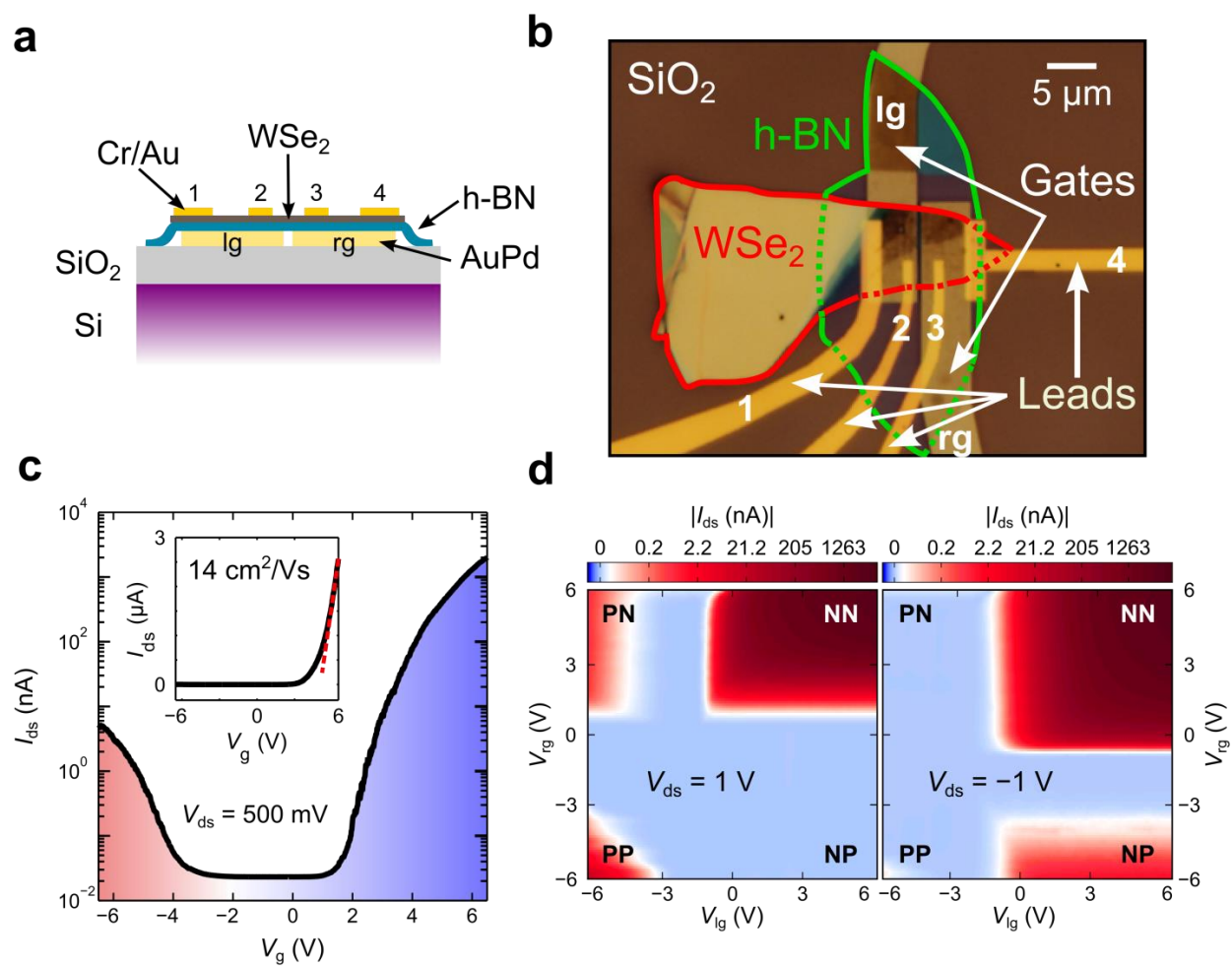

**Figure 1.** (a) Device schematic indicating the different components. (b) Optical image of the WSe$_2$ flake (outlined in red) on top of the h-BN (outlined in green) on the local gates. The leads, gates and different flakes are labeled for clarity. (c) Measured current ($I_{ds}$) as the two local gates are linked and varied as a single back-gate $V_g$, shown on a logarithmic scale. The color gradient from red to blue highlights the transition from hole-doping to electron-doping as the gates are varied from negative to positive voltages. The inset shows the data on a linear scale. (d) Color maps of the magnitude of $I_{ds}$ (on a logarithmic scale) as the voltages on the local gates are independently varied from -6 V to 6 V with a fixed positive (left) and negative (right) $V_{ds}$.

The double-gated WSe$_2$ device is measured in dark and in vacuum in an FET configuration, i.e., the local gates are linked and the voltages are varied simultaneously (acting as a single back-gate $V_g$). In Figure 1c the measured source-drain current ($I_{ds}$) is shown as $V_g$ is varied from -6 V to 6 V, while applying a fixed source-drain voltage $V_{ds}$ = 500 mV. The inset shows the same data on a linear scale. The current is strongly modulated when $V_g$ is varied from negative to positive values, showing that both hole- and electron-doping can be readily accessed due to the ambipolarity of WSe$_2$.[9,12] The threshold voltage for electron doping is lower than for hole doping indicating that the device is more n-type, as previously observed for WSe$_2$ FETs on h-BN.[12] The device displays an n-type field-effect mobility of 14.2 cm$^2$/V·s and an on/off ratio of almost 10$^5$. For p-type conduction, the device shows a mobility of 0.02 cm$^2$/V·s and an on/off ratio of 10$^2$. Note that the mobility values have been obtained from two-terminal measurements and thus they should be considered as a lower bound for the mobility as the contact resistance has not been accounted for.

By applying different voltages to the local gates we can independently control the carrier type and density on the left and right side of the flake. Figure 1d shows two color maps of the magnitude of $I_{ds}$ (on a logarithmic scale) as the left and right local gate voltages ($V_{lg}$ and $V_{rg}$ respectively) are varied independently from -6 V to 6 V, while fixing a positive (left panel) or negative (right panel) $V_{ds}$. Similar to Baugher *et al*.[10], we find that for $V_{ds}$ = 1 V a large current flows in three corners of the map; in the PP region ($V_{lg}$, $V_{rg}$ < 0 ), NN region ($V_{lg}$, $V_{rg}$ > 0) and in the PN region ($V_{lg}$ < 0, $V_{rg}$ > 0). A small current is measured in the NP region ($V_{lg}$ > 0, $V_{rg}$ < 0). For $V_{ds}$ = -1 V, we measure a large current in the NP region, whereas a small current is measured in the PN region. This indicates that when the local gates are oppositely biased (PN or NP), the *IV* curves are rectifying. The *IV* curves in different gate configurations are shown in Figure S4.

Note that the PP region shows suppressed current for negative bias, whereas there is higher current in the NN region. This is due to slightly nonlinear *IV* characteristics (Figure S5), which we ascribe to a small difference in Schottky barrier height for electrons and holes at the Cr/WSe$_2$ interface. Moreover, for both positive and negative $V_{ds}$ the current in the PP and NN configurations is three orders of magnitude larger than in the PN and NP configurations, indicating that the rectifying *IV* characteristics result from transport through the PN junction and do not arise due to the Schottky barriers. To extract the diode parameters, we fit the rectifying *IV* curves to a modified version of the Shockley equation (Equation 2 in Section 3 of the Supporting Information) which includes parasitic resistances in parallel ($R_p$) and in series ($R_s$) to the junction. The data is well fitted by the model (Figure S6) and we extract saturation currents $I_s$ in the order of 0.1 – 1 fA. All diode parameters are listed in Table S1.

Next we study the photoresponse of the gate-defined PN junction by measuring *IV* characteristics under continuous illumination in vacuum. The entire device area is illuminated by a laser beam with a diameter of 230 μm. Figure 2a shows *IV* characteristics of the device in PN configuration ($V_{lg}$ = -6 V, $V_{rg}$ = 3 V) in dark (solid line) and under illumination (dashed line). Under above-bandgap illumination (λ = 640 nm, $P$ = 4.7 W/cm$^2$) we measure a short-circuit current ($I_{sc}$, current at zero bias) of |$I_{sc}$| = 0.41 nA and an open-circuit voltage ($V_{oc}$, voltage without current flow) of $V_{oc}$ = 700 mV. The inset shows *IV* characteristics in NP configuration ($V_{lg}$ = 3V, $V_{rg}$ = -6V) in dark (solid line) and under illumination (dashed line). The increase of the slope of the *IV* curves for $V_{ds} < 0$ as a function of power indicates a photoconductive effect (an analysis of this effect can be found in the Supporting Information). Figure 2b displays *IV* characteristics in PN configuration under above-bandgap illumination (λ = 640 nm) with power densities up to 4.8 W/cm$^2$. Both $I_{sc}$ and $V_{oc}$ increase as the laser power increases. In the inset the magnitude of the generated electrical power $P_{el} = V_{ds} \cdot I_{ds}$ is plotted against $V_{ds}$; for the highest illumination power, $P_{el}$ reaches 170 pW.

In order to calculate quantities as the power conversion efficiency ($\eta_{PV}$), responsivity (*R*) and external quantum efficiency (EQE) one needs to normalize the incident power by the active area of the device. In recent works, two different conventions have been used: assuming that the entire area of the semiconducting channel is the active area[10] or considering that the active area is the region where the built-in electric field is generated by the PN junction[11]. The second convention is more realistic, but as a good estimate of the lateral size of the PN junction is difficult to achieve, we consider the active area to be the entire semiconducting channel region between the source and drain. Hence, the power conversion efficiency, responsivity and external quantum efficiency values reported here are very conservative lower bounds. The power conversion efficiency is defined as $\eta_{PV} = P_{el}/P_{in}$, where $P_{in}$ is the power incident on the active area of the device. For the highest incident power in Figure 2b we calculate $\eta_{PV} \approx 0.01\%$. A higher value (≈ 0.5%) has previously been reported, however here it was assumed that that the power conversion takes place in the intrinsic (ungated) device region.[11] Using the same definition, we obtain a value of 0.14%.

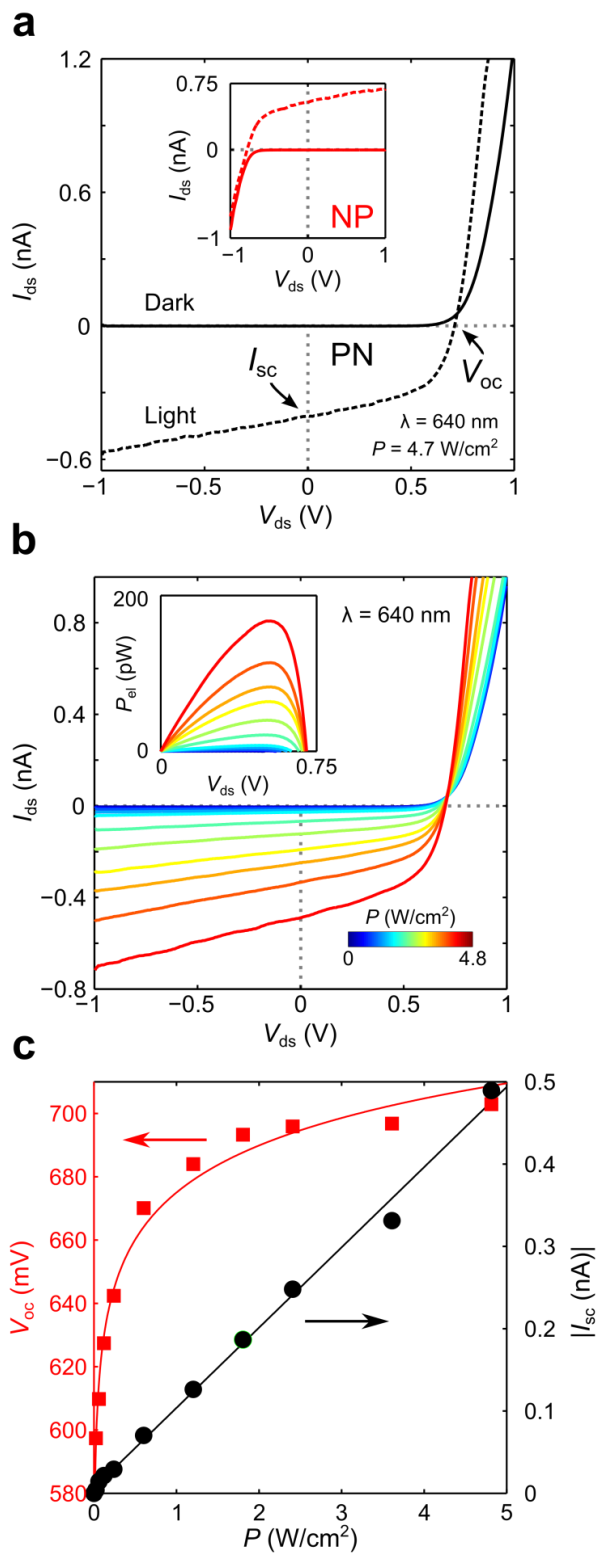

**Figure 2.** (a) *IV* characteristics in PN (black) and NP (red, inset) configurations, where the local gates are oppositely biased (PN: $V_{lg}$ = -6 V, $V_{rg}$ = 3 V; NP: $V_{lg}$ = 3 V, $V_{rg}$ = -6 V). Data is shown for measurements in dark (solid lines) and under 640 nm illumination (dashed lines) with $P$ = 4.8 W/cm$^2$. (b) *IV* characteristics in PN configuration under illumination with different powers incident $P$ on the device, ranging up to 4.8 W/cm$^2$. The inset shows the electrical power $P_{el}$ generated by the device, calculated as $P_{el} = V_{ds} \cdot I_{ds}$. The maximum electrical power generated is around 170 pW. (c) Open-circuit voltage ($V_{oc}$, left axis, red squares) and short-circuit current ($I_{sc}$, right axis, black circles) against power density $P$, extracted from the data in (b). $I_{sc}$ follows a linear dependence on the power (linear fit, black line), whereas $V_{oc}$ follows a logarithmic dependence (logarithmic fit, red line).

In an ideal solar cell, $I_{sc}$ depends linearly on the illumination power density ($P$) and $V_{oc}$ depends logarithmically on $P$ due to the exponential shape of the *IV* characteristic. In Figure 2c, $V_{oc}$ (left axis, red squares) and $I_{sc}$ (right axis, blue circles) are plotted against $P$. The good agreement with the fitted curves (black line, linear fit to $I_{sc}$; red line, logarithmic fit to $V_{oc}$) confirms that the photocurrent generation in PN configuration is dominated by the photovoltaic effect. Since $I_{sc}$ is linearly dependent on $P$, we extract the responsivity $R = I_{sc}/P_{in}$. This yields $R = 0.24$ mA/W, which is the responsivity of the PN junction at zero bias under 640 nm illumination.

The performance of our device, both in dark and under different illumination powers, is comparable to that of recently reported WSe$_2$ PN junctions.[10-12] We further characterize the PN junction by studying the photoresponse under different illumination wavelengths and by measuring the response time using modulated light excitation. Figure 3a shows *IV* characteristics of the device in PN configuration under different illumination wavelengths. The *IV* characteristics measured with 940 nm, 885 nm and 808 nm wavelengths show no photoresponse, coinciding with the trace in dark. From the *IV* characteristics we extract the responsivities at $V_{ds} = -1$ V and $V_{ds} = 0$ V, which show a maximum at $\lambda = 532$ nm and a cut-off wavelength of $770 \pm 35$ nm, consistent with the bandgap of single-layer WSe$_2$ (1.65 eV, 750 nm).[19] Here, the cut-off wavelength is defined as the wavelength above which no photocurrent is measured when the device is biased in PN configuration. The maximum responsivity value measured at $V_{ds} = -1$ V is 0.70 mA/W under 532 nm illumination. This value is lower than previously reported values for WSe$_2$ PN junctions, likely due to a lower applied source-drain voltage in our device or due to a different definition of the active area.[10,11] From the data in Figure 3a we calculate the external quantum efficiency (EQE) values, given by $EQE = I_{sc}/P_{in} \cdot (hc/e\lambda)$, where $h$, $c$ and $e$ are Planck's constant, the speed of light and the electron charge respectively. The EQE values as a function of wavelength are shown in Figure S8 and reach a maximum of 0.1% at 532 nm illumination.

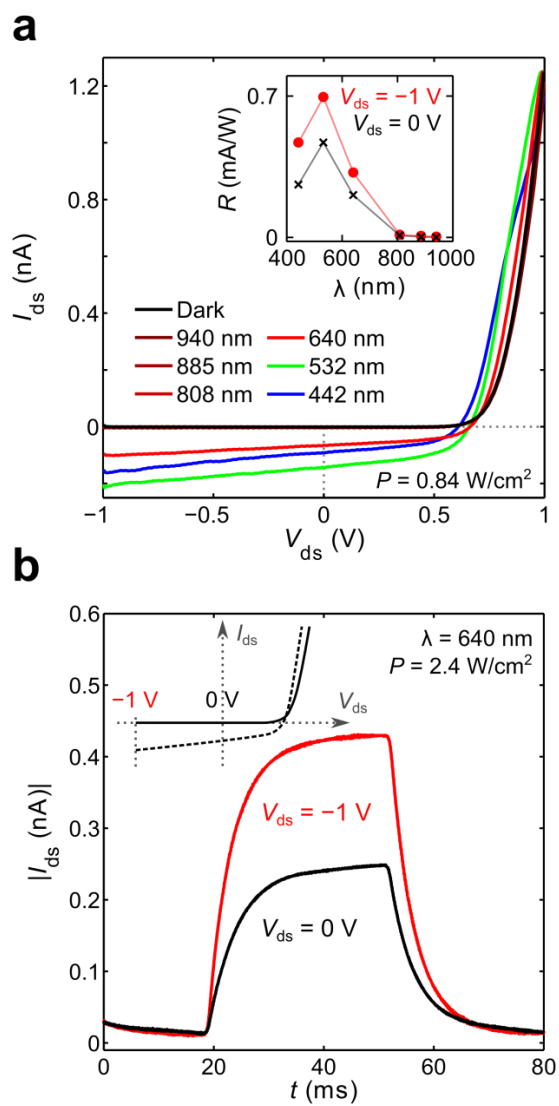

**Figure 3.** (a) *IV* characteristics in PN configuration under 442 to 940 nm illumination wavelengths. The *IV*s measured under 808 to 940 nm illumination wavelengths show no photoresponse. The inset shows a plot of the responsivity $R$ against the wavelength $\lambda$ calculated at $V_{ds} = -1$ V (red dots) and $V_{ds} = 0$ V (black crosses). (b) Magnitude of the photocurrent $I_{ds}$ as a function of time under modulated light excitation (15 Hz) under 640 nm illumination ($P = 2.4$ W/cm$^2$). $I_{ds}$ is measured as the gates are biased in PN configuration and $V_{ds}$ is fixed to 0 V (black line) or -1 V (red line).

The response time of the gate-defined PN junction to light excitation is studied by mechanically modulating the intensity of the incoming light and recording the current under constant $V_{ds}$ in the PN configuration. Figure 3b shows the current $|I_{ds}|$ against time $t$ for one period of light excitation of $\lambda$ = 640 nm and $P$ = 2.4 W/cm$^2$. The inset displays a schematic, indicating at which source-drain voltage $I_{ds}$ is measured. For zero bias (black line), we measure a rise time of $\tau_{rise}$ = 12.3 ± 0.2 ms and a fall time of $\tau_{fall}$ = 11.4 ± 0.2 ms. When the PN junction is reverse biased ($V_{ds}$ = -1 V, red line) we measure a larger current, and rise and fall times of $\tau_{rise}$ = 10.4 ± 0.1 ms and $\tau_{fall}$ = 9.8 ± 0.2 ms. The determination of $\tau_{rise}$ and $\tau_{fall}$ is explained in detail in Section 7 of the Supporting Information. To our knowledge, this is the first measurement of the response time of a PN junction based on TMDCs to date.[20-23] Hence, we compare the rise and fall times of our device to phototransistors based on single-layer MoS$_2$ and few-layer WS$_2$, as they are considered similar systems. For single-layer MoS$_2$ phototransistors the reported rise and fall times in the literature range from $\tau_{rise}$ = $\tau_{fall}$ = 50 ms to $\tau_{rise}$ = 4 s, $\tau_{fall}$ = 9 s.[6,24] Rise and fall times for a few-layer WS$_2$ phototransistor were reported to be close to 5.3 ms.[25] The response times measured in our study are thus significantly shorter than values for phototransistors based on MoS$_2$ and slightly larger than a photosensor based on WS$_2$. However, more research on response times of other electrostatically defined PN junctions is required to allow a fair comparison.

We now focus on the possible mechanisms which are responsible for the generation of photocurrent in the double-gated WSe$_2$ device. As theoretically proposed by Song *et al.*, the analysis of 2D gate maps of the photocurrent is a powerful method to identify the different photocurrent generation mechanisms in electrostatically defined PN junctions.[26] The 2D gate maps are acquired by measuring the photocurrent (or photovoltage) under illumination while the local gate voltages ($V_{lg}$, $V_{rg}$) are varied independently. The magnitude of $I_{sc}$ (or $V_{oc}$) is then represented in a colormap form with $V_{lg}$ and $V_{rg}$ on the $x$ and $y$ axis respectively.

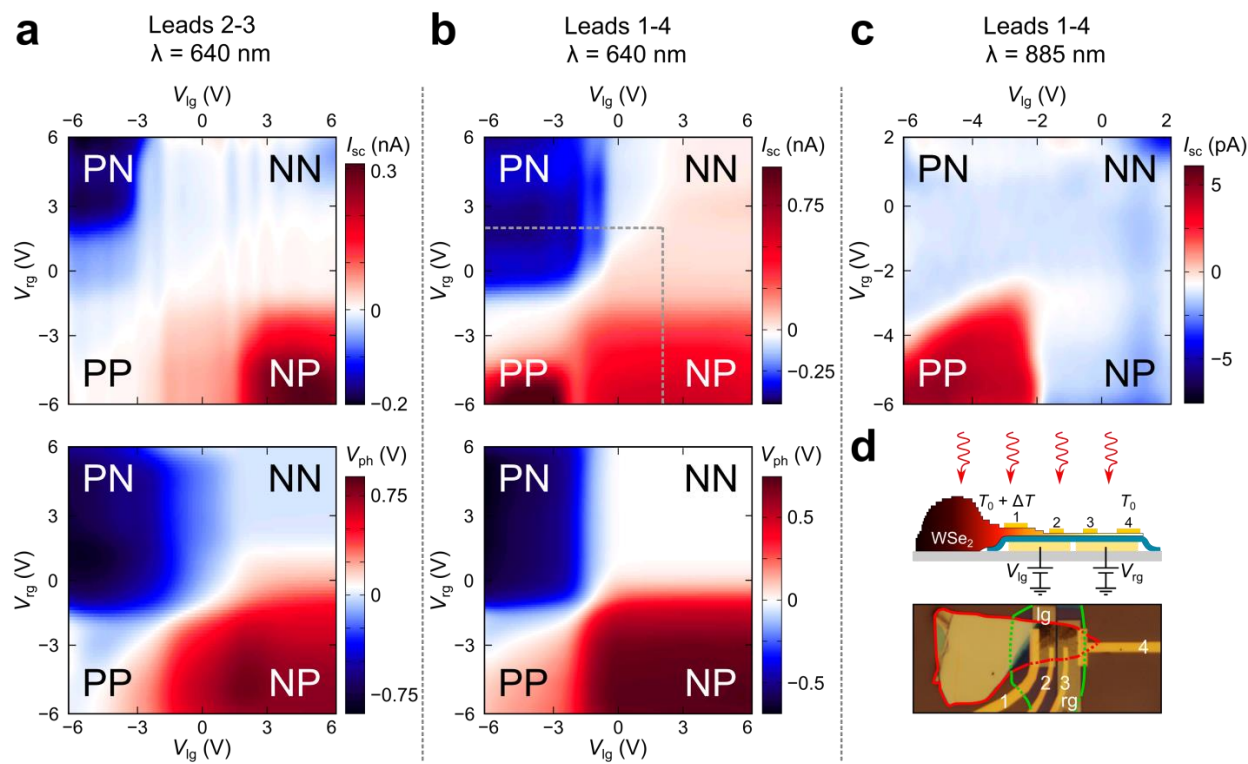

**Figure 4.** (a) 2D gate map of the photocurrent $I_{sc}$ (top) and the photovoltage $V_{oc}$ (bottom) under above-bandgap illumination ($\lambda = 640$ nm, $P = 2.4$ W/cm$^2$) as $V_{lg}$ and $V_{rg}$ are varied independently from -6 V to 6 V. The current flow is measured between leads 3 and 2 (grounded). (b) 2D gate map of $I_{sc}$ (top) and $V_{oc}$ (bottom) under above-bandgap illumination ($\lambda = 640$ nm, $P = 2.4$ W/cm$^2$) as $V_{lg}$ and $V_{rg}$ are varied independently from -6 V to 6 V. The current flow is measured between leads 4 and 1 (grounded). A 2D gate map of the current measured in dark has been subtracted, as shown in Figure S11. (c) 2D gate map of $I_{sc}$ under below-bandgap illumination ($\lambda = 885$ nm, $P = 0.72$ W/cm$^2$) as $V_{lg}$ and $V_{rg}$ are varied independently from -6 V to 2 V. The current flow is measured between leads 4 and 1 (grounded). (d) Schematic of the device, with emphasis on the thickness distribution of the flake. The laser photons (red arrows) are absorbed, heating the WSe$_2$ flake. The thicker part of the flake (left) has a stronger absorption, and hence, locally heats up. This gives rise to a temperature gradient $\Delta T$ decreasing from left to right, highlighted by the color gradient. Bottom: optical micrograph of the device, indicating the different leads and gates.

Figure 4a (top) shows a 2D gate map of the photocurrent under above-bandgap illumination ($\lambda$ = 640 nm, $P$ = 2.4 W/cm$^2$), where the photocurrent $I_{sc}$ is measured between the inner leads (3 and 2, grounded) as the local gate voltages are independently varied from -6 V to 6 V. A large $I_{sc}$ is measured in the PN and NP configurations (top left and bottom right corners, respectively) where the two gates are oppositely biased, caused by the separation of photoexcited electron-hole pairs by the built-in electric field of the PN junction. The generation of a short-circuit current from the photovoltaic effect is associated with an open-circuit voltage; this is reflected in Figure 4a (bottom), where a 2D gate map of $V_{oc}$ under above-bandgap illumination is shown in the same gate voltage range. As expected, a large open-circuit voltage is measured in the PN and NP configurations. The large $I_{sc}$ and $V_{oc}$ generated in the PN and NP configurations confirm that the dominant photocurrent generation mechanism is the photovoltaic effect and the photocurrent does not arise from Schottky barriers at the contacts, as photocurrent generated by Schottky barriers would also be present in the PP and NN regions. In fact, as we employ an extended spot (illuminating the complete device) one would require strongly asymmetric Schottky barriers at the contacts (which usually involves the use of different contact metals for the source and drain electrodes) to generate Schottky barrier driven photocurrent.[27]

Instead of measuring between the inner leads, which are closer spaced and only contact the monolayer part of the WSe$_2$ flake, we measure the current flow between leads 4 and 1 (grounded). These leads define a longer channel and contact different thicknesses of the WSe$_2$ flake. In Figure 4b (top) a 2D gate map of $I_{sc}$ under above-bandgap illumination is shown, where $V_{lg}$ and $V_{rg}$ are varied from -6 V to 6 V. Again, we measure a large $I_{sc}$ (0.5 nA) in the PN and NP configurations due to the photovoltaic effect. Interestingly, we also measure a large photocurrent (1.1 nA) in the PP region. In Figure 4b (bottom) a 2D gate map of $V_{oc}$ under above-bandgap illumination is shown, where a large photovoltage is present in the PN and NP configurations (700 mV) arising from the photovoltaic effect. A smaller photovoltage (130 mV) is measured in the PP region. To exclude separation of photoexcited electron-hole pairs as the origin of the this photocurrent and photovoltage, we measure $I_{sc}$ in combination with below-bandgap illumination. A 2D gate map of the photocurrent under below-bandgap illumination ($\lambda$ = 885 nm, $P$ = 0.72 W/cm$^2$) is shown in Figure 4c, where $V_{lg}$ and $V_{rg}$ are varied from -6 V to 2 V. As expected, the photocurrent in the PN and NP configurations disappears. However, the photocurrent in the PP region is still present. The persistence of this region of photocurrent under below-bandgap illumination and the small photovoltage generated suggest that this photocurrent originates from the photothermoelectric effect, as previously observed in single-layer MoS$_2$.[16] In the photothermoelectric effect, a temperature gradient $\Delta T$ arising from light absorption generates a photothermal voltage across two junctions between two materials with different Seebeck coefficient $S$. The photothermal voltage, given by $\Delta V_{PTE} = \Delta S \cdot \Delta T$, can drive a current through the device.

The emergence of the photothermoelectric effect in our device can be attributed to the presence of a temperature gradient due to a region of strong light absorption in the flake. As shown in Figure 4d, the WSe$_2$ flake consists of a thick part (50 nm) attached to a thin part. Since absorption increases as a function of flake thickness and the entire device area is illuminated, the thick part of the flake is locally heated, resulting in a temperature gradient highlighted by the color gradient in Figure 4d (top). Note that the sign of the photothermoelectric current is dictated by the sign of the Seebeck coefficient (positive for p-type semiconductors and negative for n-type, $S$ of the metal contacts is negligible) and the direction of the thermal gradient. The observed photocurrent is consistent with a p-type semiconductor and a temperature gradient from lead 1 (hot) to lead 4 (cold). A photothermoelectric current is also present, although smaller, when measuring under below-bandgap illumination between the inner leads (~ 1 pA), as these leads are closely spaced and thus $\Delta T$ is smaller (Figure S10). A possible explanation for the absence of photocurrent in the NN configuration is a smaller Seebeck coefficient for n-doped $WSe_2$, as expected by the low device resistance in the NN configuration. However, since a reduction in resistance can also arise from smaller Schottky barriers (due to field-effect or illumination), a detailed study including four-terminal measurements is required to determine the ratio between the Seebeck coefficients of p- and n-doped $WSe_2$. We attribute the absence of photothermoelectric current in the PN and NP configurations to the large tunnel resistance at the depletion region of the PN junction. When measuring between leads 1 and 4, the current generated by the photothermoelectric effect (1.1 nA) in the PP configuration is more than a factor two larger than the current generated by the photovoltaic effect (0.5 nA) in the PN and NP configurations. Nonetheless, the high absorbed optical power and small generated thermal voltage render the photothermoelectric effect approximately $1.5 \cdot 10^4$ times less efficient for power generation than the photovoltaic effect in the device (Section 10 of the Supporting Information).

Hence, we find that both photovoltaic and photothermoelectric effects can generate photocurrent in double-gated WSe2 devices. 2D gate maps with above-bandgap illumination show these two effects: current due to the photovoltaic effect is present in the PN and NP configurations and current of photothermoelectric origin appears in the PP configuration. As expected, the magnitude of the photothermoelectric current is smaller in the short channel $WSe_2$ (small $\varDelta T$) than in the long channel (large $\varDelta T$). When measuring with below-bandgap illumination (Figure 4c) we do not observe current generated by the photovoltaic effect and only the current of photothermoelectric origin remains. Hence, the analysis of 2D gate maps of the photocurrent (and photovoltage) in combination with above- and below-bandgap illumination provides a powerful method to disentangle the photovoltaic and photothermoelectric effect.

In conclusion, we have fabricated an electrostatically tunable device based on single-layer $WSe_2$ whose output can be controlled by means of local gating. Oppositely biasing the gates causes the formation of a PN junction in the channel, displaying rectifying *IV* characteristics. Under above-bandgap illumination, a large photocurrent (and photovoltage) is generated in the PN and NP configurations by the photovoltaic effect. The device displays short-circuit currents up to 0.5 nA and open-circuit voltages up to 700 mV. A maximum electrical power of 170 pW is generated by the device. Transport studies as a function of illumination power show that the photocurrent generation in the PN/NP configurations is dominated by the photovoltaic effect. These results are comparable to recent reports on electrostatically defined $WSe_2$ PN junctions with similar geometries. In addition, we characterized the photoresponse under different illumination wavelengths and found a maximum responsivity of 0.70 mA/W at 532 nm and a cut-off wavelength of 770 ± 35 nm, in agreement with the bandgap of single-layer $WSe_2$. Moreover, the response time of the PN junction under modulated light excitation was measured at zero bias and in reverse bias. We determined rise and fall times close to 10 ms, significantly shorter than for phototransistors based on $MoS_2$. Finally, 2D gate maps of the photocurrent and photovoltage under above- and below bandgap illumination enabled us to observe and disentangle two different main photocurrent generation mechanisms: the photovoltaic and photothermoelectric effect. Our results demonstrate that the photothermoelectric effect can play a significant role in $WSe_2$ devices where a region of strong optical absorption generates a sizeable thermal gradient under illumination.

METHODS

*Fabrication of double-gated WSe₂ device.* The double-gated $WSe_2$ devices with hexagonal boron nitride (h-BN) as gate dielectric and tungsten diselenide ($WSe_2$) as channel material was fabricated using an all-dry transfer technique, as described in Refs. [18,29] We start by fabricating a pair of local gates (Ti/AuPd, 5 nm/25 nm) with a separation of 300 nm on an $SiO_2$ (285 nm)/Si substrate using e-beam lithography (*Vistec, EBPG5000PLUS HR 100*), metal deposition (*AJA international*) and subsequent lift-off (warm acetone). Then, h-BN is exfoliated from powder (Momentive, Polartherm grade PT110) onto a viscoelastic stamp (GelFilm® by GelPak) with blue Nitto tape (Nitto Denko Co., SPV 224P). We identify thin and homogeneous h-BN flakes by optical inspection and transfer them onto the local gates by pressing the viscoelastic stamp against the substrate and slowly releasing it. We then exfoliate $WSe_2$ from a synthetic crystal grown by vapor transport method and identify a single-layer flake and transfer it on top of the h-BN flake using the same method. Contact to the flake is made by patterning leads (Cr/Au, 0.3 nm/60 nm) by e-beam lithography, metal deposition and lift-off.

*Characterization of h-BN and WSe$_2$*. Atomic Force Microscopy (AFM) is used to determine the thickness of the h-BN and WSe$_2$ flakes. The AFM (*Digital Instruments D3100 AFM*) is operated in amplitude modulation mode with Silicon cantilevers (spring constant 40 N m$^{-1}$ and tip curvature <10 nm). Raman spectroscopy (Renishaw in via) was performed in a backscattering configuration excited with a visible laser light (λ = 514 nm). Spectra were collected through a 100× objective and recorded with 1800 lines/mm grating providing a spectral resolution of ~ 1 cm$^{-1}$. To avoid laser-induced heating and ablation of the samples, all spectra were recorded at low power levels P ~ 500 μW and short integration times (1 sec). Photoluminescence measurements have been carried out with the same Renishaw in via setup but longer integration times were needed (60-180 sec).

*Optoelectronic characterization.* Optoelectronic characterization is performed in a *Lakeshore Cryogenics* probe station at room temperature in vacuum (~ 10$^{-5}$ mbar). The light excitation is provided by diode pumped solid state lasers operated in continuous wave mode (CNI Lasers). The light is coupled into a multimode optical fiber (NA = 0.23) through a parabolic mirror ($f_{\text{reflected}}$ = 25.4 mm). At the end of the optical fiber, an identical parabolic mirror collimates the light exiting the fiber. The beam is then directed into the probe station. The beam spot size on the sample has a diameter of 230 ± 8 μm for all used wavelengths. The powers mentioned in the main text are the power densities in W/cm$^2$, calculated by $P = P_{\text{laser}}/A_{\text{spot}}$, where $P_{\text{laser}}$ is the laser power and $A_{\text{spot}}$ the area of the laser spot. A 10 MΩ resistor was used to characterize the response time of the measurement setup (0.2 ms). The photovoltage measurements were performed by directly measuring the open-circuit voltage with respect to ground.

ACKNOWLEDGEMENT

This work was supported by the Dutch organization for Fundamental Research on Matter (FOM). A.C-G. acknowledges financial support through the European Union FP7-Marie Curie Project PIEF-GA-2011-300802 ('STRENGTHNANO').

Supporting Information for

# Photovoltaic and Photothermoelectric Effect in a Double-Gated WSe$_2$ Device

*Dirk J. Groenendijk*[*,†], *Michele Buscema*[†], *Gary A. Steele*[†], *Steffen Michaelis de Vasconcellos*[‡], *Rudolf Bratschitsch*[‡], *Herre S.J. van der Zant*[†] *and Andres Castellanos-Gomez*[*,†,+]

[†] Kavli Institute of Nanoscience, Delft University of Technology, Lorentzweg 1, 2628 CJ Delft (The Netherlands).
[‡] Institute of Physics, University of Münster, D-48149 Münster (Germany).
[+] Present address: Instituto Madrileño de Estudios Avanzados en Nanociencia (IMDEA-Nanociencia), 28049 Madrid (Spain)

**Table of contents**

**1. Device fabrication - optical images of the transfer steps**

The fabrication steps for the double-gated $WSe_2$ device are presented in Figure S1 and consist of:

1. Fabrication of local gates (Ti/AuPd, 5 nm/25 nm) with a separation of 300 nm on a $Si/SiO_2$ substrate (Figure S1a);

2. Exfoliation and deterministic transfer of a thin (10 nm) h-BN flake as gate dielectric (Figure S1b);

3. Exfoliation and deterministic transfer of a $WSe_2$ flake as channel material (Figure S1c);

4. Fabrication of leads (Cr/Au, 0.3 nm/60 nm) to make contact to $WSe_2$ (Figure S1d).

Optical images of the h-BN and $WSe_2$ flake on the viscoelastic stamp prior to transfer are shown in Figure S1e and f respectively.

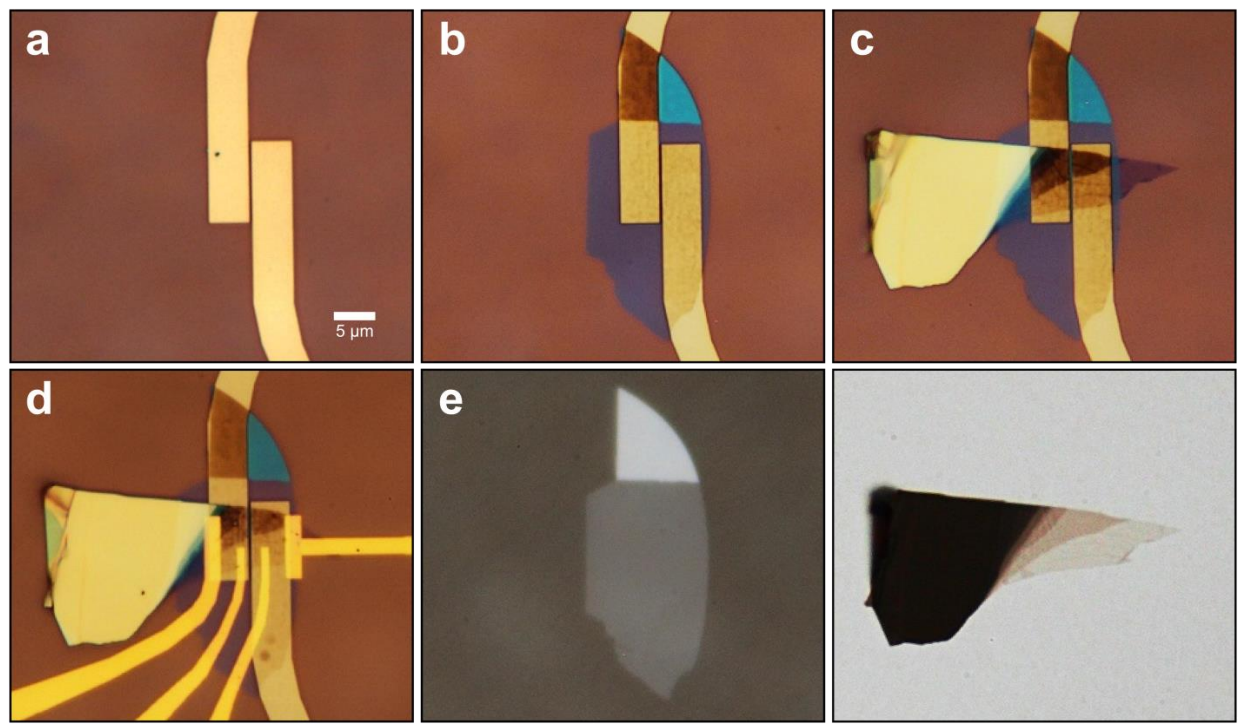

**Figure S1.** (a) Local gates (AuPd) patterned on a Si/ SiO$_2$ substrate. The gap between the gates is 300 nm wide. (b) A thin (10 nm) h-BN flake situated on top of the local gates. **(c)** A thin WSe$_2$ flake on top of the h-BN and the split gates. (d) The final device after fabrication of the contacts (Cr/Au). (e) Optical image (reflection mode) of the h-BN flake on the viscoelastic stamp. (f) Optical image (transmission mode) of the WSe$_2$ flake on the viscoelastic stamp.

## 2. Characterization of the 2D materials (AFM, PL, Raman)

An AFM topographic image of the device is shown in Figure S2.

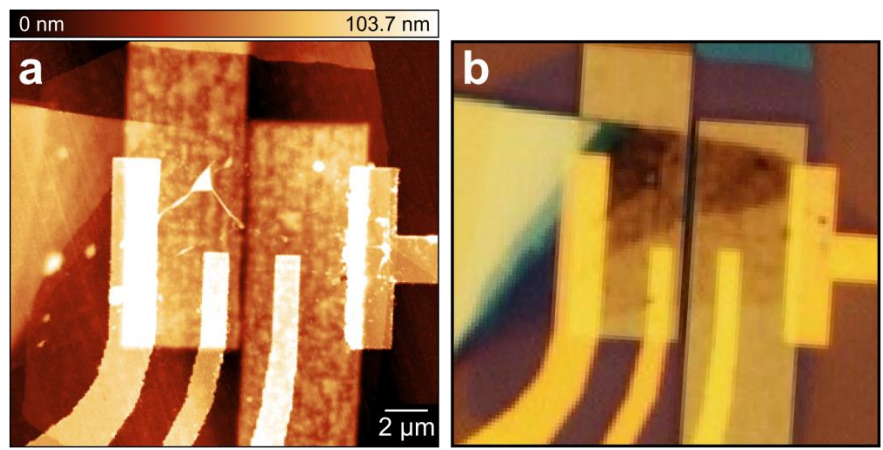

**Figure S2.** (a) Atomic Force Microscopy (AFM) topographic image of the double-gated WSe$_2$ device. (b) Corresponding optical image.

Photoluminescence and Raman spectroscopy measurements confirm the single-layer thickness of the WSe$_2$ flake in the region within the inner leads, as shown in Figure S3.

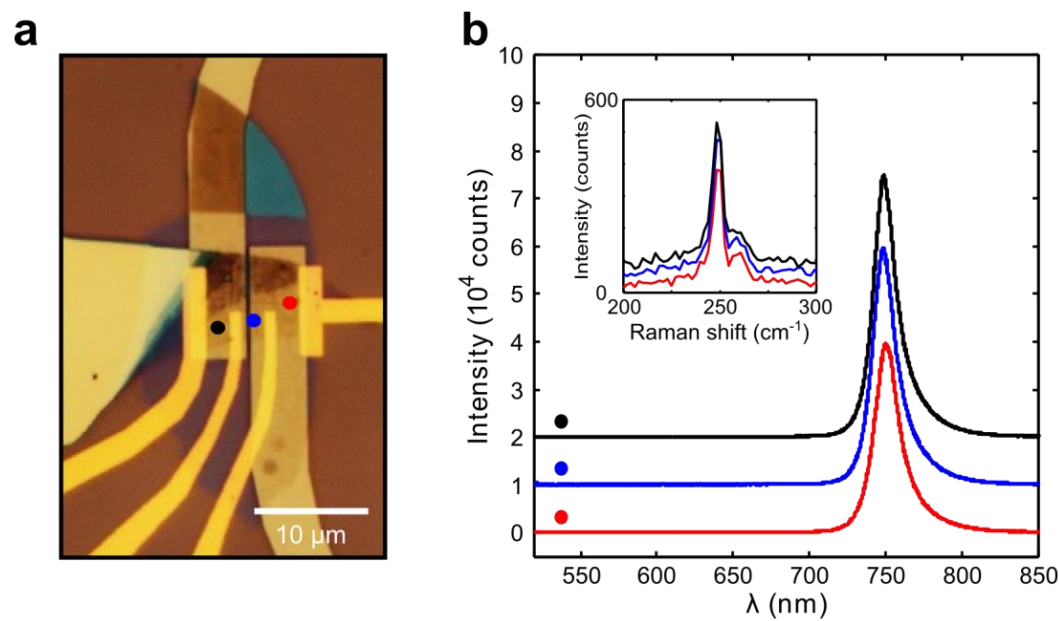

**Figure S3.** (a) Optical image of the WSe$_2$ flake on the h-BN. The spots indicate the positions where the different measurements are taken. (b) Photoluminescence measurements, showing the distinct peak of single-layer WSe$_2$ around 750 nm (1.65 eV). The spectra have been shifted vertically by $10^4$ counts for clarity. The inset shows the result of the Raman measurements.

### 3. Output characteristics of the device

The *IV* characteristics of the device in different gate configurations is shown in Figure S4. The device can be operated as a resistor (PP and NN configurations) or as a diode in either direction (PN and NP configurations).

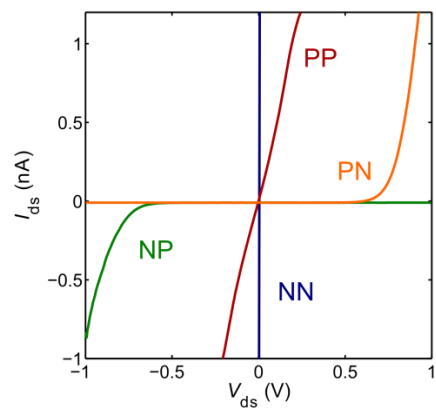

**Figure S4.** *IV* characteristics in different gate configurations (PP: $V_{lg} = V_{rg} = -6$ V, NN: $V_{lg} = V_{rg} = 6$ V, PN: $V_{lg} = -6$ V, $V_{rg} = -3$ V, NP: $V_{lg} = 3$ V, $V_{rg} = -6$ V).

*IV* characteristics at different linked local gate voltages are shown in Figure S5.

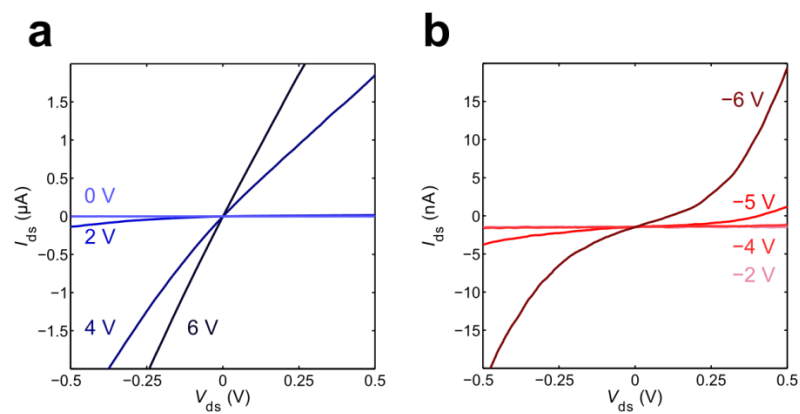

**Figure S5.** (a) *IV* characteristics for $V_g = 0$ V to 6 V. (b) *IV* characteristics for $V_g = -6$ V to -2 V.

## 4. Extraction of the diode parameters

The Shockley equation can be modified to include resistances in parallel ($R_p$) and in series ($R_s$) with the junction. Its modified form is given by[1]

$$I_{ds} = I_s \cdot \left[\exp\left(\frac{V_{ds} - I_{ds}R_s}{nV_T}\right) - 1\right] + \frac{V_{ds} - I_{ds}R_s}{R_p}. \quad (1)$$

An explicit solution for $I_{ds}$ can be written in terms of the Lambert W-function:

$$I_{ds} = \frac{nV_T}{R_s} \mathrm{W}\left\{\frac{I_s R_s R_p}{nV_T(R_s + R_p)} \exp\left(\frac{R_p(V_{ds} + I_s R_s)}{nV_T(R_s + R_p)}\right)\right\} + \frac{V_{ds} - I_s R_p}{R_s + R_p}, \quad (2)$$

which is used to fit the *IV* curves in PN and NP configurations. The result of the fits is shown in Figure S6.

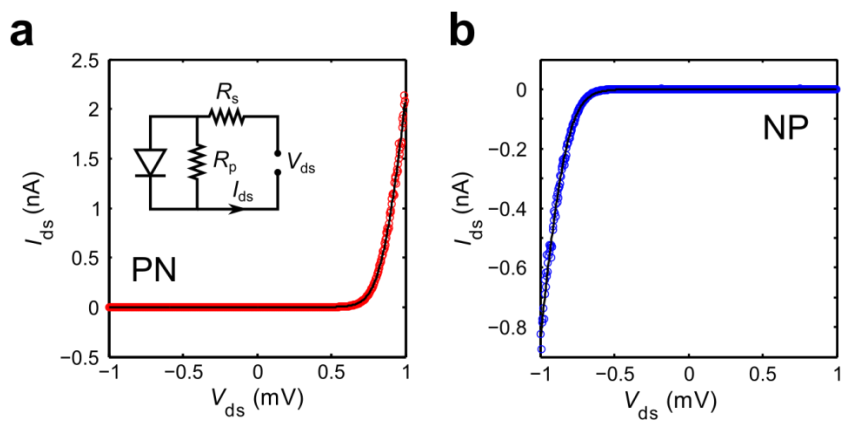

**Figure S6.** (a) *IV* curve in PN configuration (red circles) fitted by the model (black line). The inset shows the circuit model. (b) *IV* curve in NP configuration (blue circles) fitted by the model (black line).

The diode parameters extracted from the fits are listed in Table S1. The values compare well to those mentioned in previous studies.[2,3]

|  | PN | NP |
|---|---|---|
| $I_s$ (fA) | 0.19 | 1.10 |
| $n$ | 2.14 | 2.56 |
| $R_s$ (MΩ) | 49.4 | 122.0 |
| $R_p$ (MΩ) | ∞ | ∞ |

**Table S1.** Extracted diode parameters.

### 5. Photoconductance in reverse bias in PN configuration

The *IV* curves in Figure 2b can be fitted to a linear function in the region $V_{ds} < 0$, enabling us to extract a resistance R as a function of illumination power (Figure S7). *R* decreases as *P* increases, as shown in Figure S7b. As the resistance decreases with increasing *P*, it can be associated with photoconductivity of the WSe$_2$ channel.

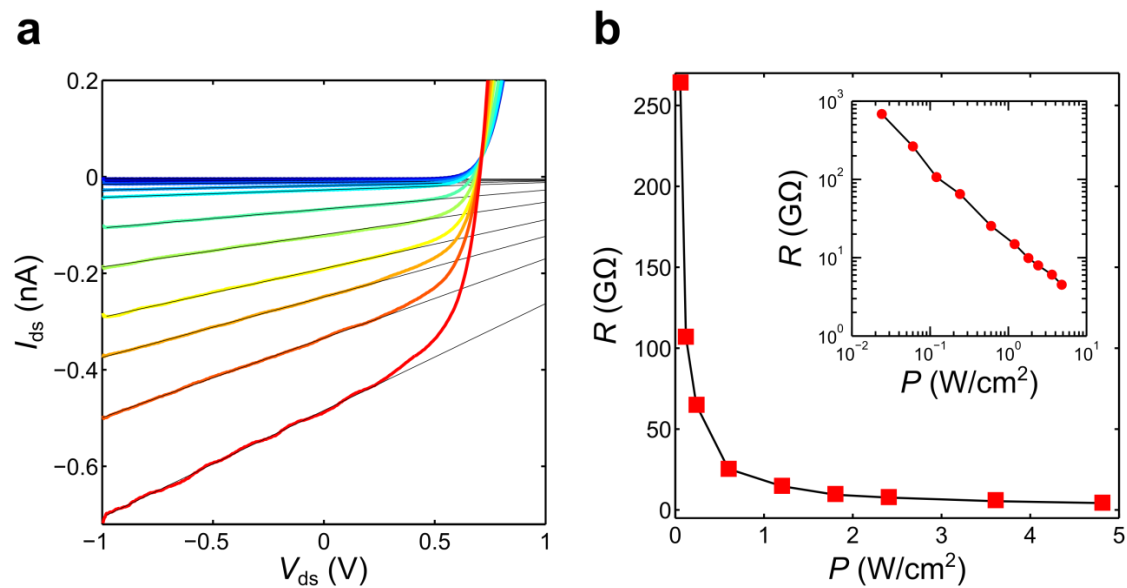

**Figure S7.** (a) IV characteristics in PN configuration ($V_{lg}$ = -6 V, $V_{rg}$ = 3 V) under illumination with different powers. The black lines are linear fits to the current measured between $V_{ds}$ = -1 V and 0 V. (b) Resistance values as a function of power, extracted from the fits. The inset shows the data on a log-log scale.

## 6. External quantum efficiency (EQE) as a function of wavelength

From the data in Figure 3a we calculate the EQE values as a function of wavelength, which is presented in Figure S9.

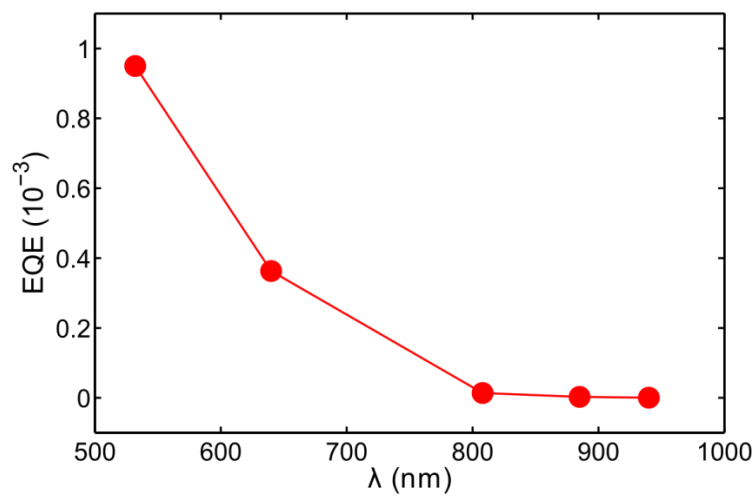

**Figure S8.** EQE as a function of wavelength for the gate-defined WSe$_2$ PN junction. The EQE reaches a maximum of about 0.1% at 532 nm illumination.

## 7. Determination of response times

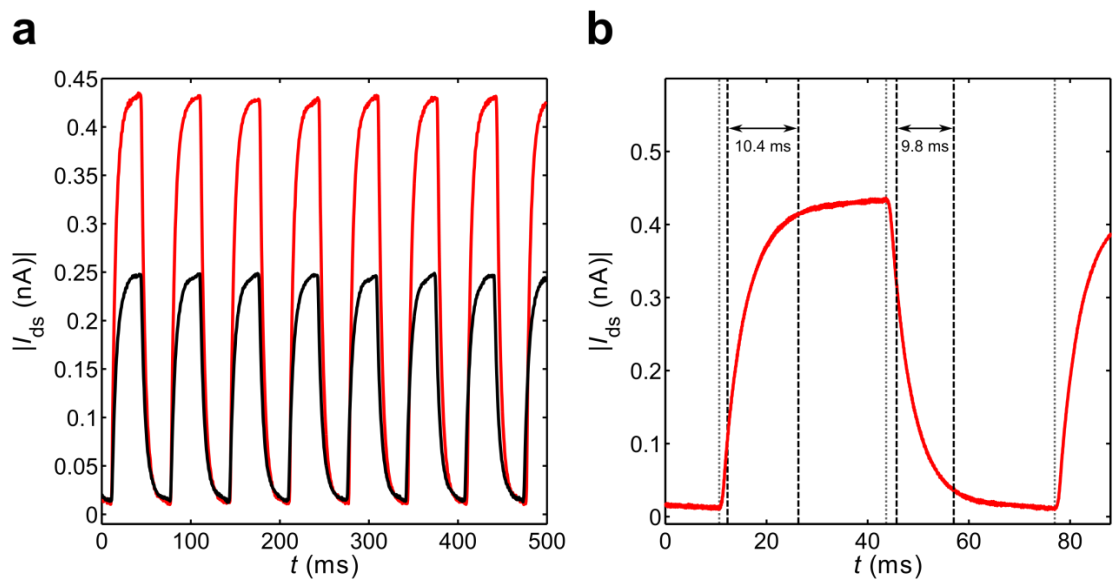

**Figure S9.** (a) Magnitude of the photocurrent $I_{ds}$ as a function of time under modulated light excitation (~ 15 Hz) under 640 nm illumination ($P = 2.4$ W/cm$^2$). $I_{ds}$ is measured as the gates are biased in PN configuration and $V_{ds}$ is fixed to 0 V (black line) or -1 V (red line). Seven periods are shown. (b) One period of modulated light excitation for $V_{ds} = -1$ V. Lines are drawn to indicate the times where the rise and fall times $\tau_{rise}$ and $\tau_{fall}$ are calculated according to the 10%-90% criterion.

The rise and fall times $\tau_{rise}$ and $\tau_{fall}$ are extracted from seven periods of light excitation (Figure S9a) in order to determine uncertainties. As shown in Figure S5b, the values are the 10%-90% rise and fall times. The values for all seven periods are shown in Table S1. Due to the high resistance of the WSe$_2$ device (170 MΩ), the response time is limited by the *RC* bandwidth. We note that the response time of our measurement setup is around 0.2 ms ($f_{3db} = 700$ Hz), measured across a 10 MΩ resistor.

| $V_{ds} = 0$ V | | $V_{ds} = -1$ V | |
|---|---|---|---|
| $\tau_{rise}$ (ms) | $\tau_{fall}$ (ms) | $\tau_{rise}$ (ms) | $\tau_{fall}$ (ms) |
| 12.71 | 11.25 | 10.33 | 9.92 |
| 12.18 | 11.65 | 10.44 | 9.60 |
| 12.08 | 11.46 | 10.20 | 9.80 |
| 12.34 | 11.26 | 10.48 | 9.67 |
| 12.07 | 11.36 | 10.61 | 9.78 |
| 12.13 | 11.25 | 10.55 | 10.08 |
| 12.36 | 11.62 | 10.47 | 9.84 |

**Table S2.** Rise and fall times for seven periods of light excitation.

## 8. Below-bandgap current measured between the inner leads

A 2D gate map of the photocurrent under below-bandgap illumination (λ = 885 nm, P = 0.72 W/cm$^2$) is shown in Figure S10. We measure a small (1 pA) photocurrent in the PP configuration and no current in the PN, NP and NN configurations. The magnitude of the photocurrent measured between the inner leads is 3 orders of magnitude smaller than that measured between the outer leads due to the difference in channel length and $\Delta T$.

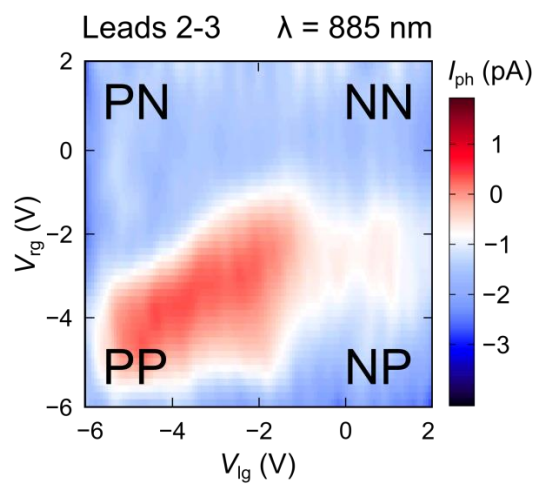

**Figure S10.** 2D gate map of the photocurrent generated under below-bandgap illumination measured between leads 3 and 2 (grounded).

### 9. Subtraction of a 2D gate map of the current in dark

Due to leakage for positive gate voltages when measuring between the outer leads, a 2D gate map of the current measured in dark has been subtracted, as shown in Figure S11.

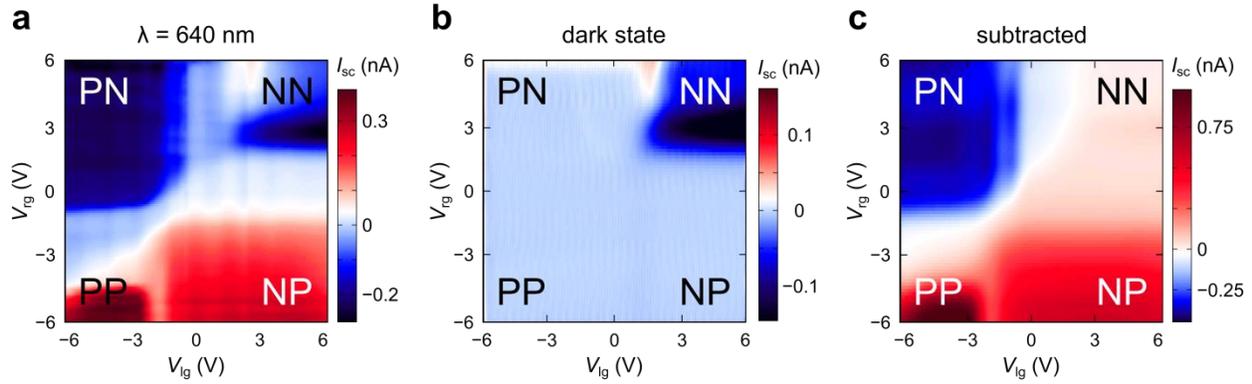

**Figure S11.** (a) 2D gate map of $I_{sc}$ before subtraction of the current in dark under above-bandgap illumination ($\lambda$ = 640 nm, P = 2.4 W/cm$^2$) as the voltages on the two local gates are varied independently from -6 V to 6 V. The current flow is measured between leads 4 and 1 (grounded). (b) 2D gate map of $I_{sc}$ in dark in the same voltage range as in panel (a), measured between leads 4 and 1 (grounded). (c) 2D gate map of $I_{sc}$ after subtraction of the 2D gate map in dark.

## 10. Photovoltaic and photothermoelectric power conversion efficiencies

For a quantitative comparison of the power generation by the photovoltaic and the photothermoelectric effect, we consider the maximum power conversion efficiencies $\eta_{PV,max}$ and $\eta_{PTE,max}$ (neglecting losses) based on the data in Figure 4b. Here, $\eta_{max}$ is defined as

$$\eta_{max} = P_{el,max}/P_{in},$$

where $P_{el,max} = I_{sc} \times V_{oc}$ is the maximum obtainable generated power and $P_{in}$ is the absorbed optical power. For the photovoltaic effect, we assume the power conversion takes place in the single-layer part of the flake, whereas for the photothermoelectric effect, we assume power conversion takes place in the (50 nm) thick part of the WSe$_2$ flake. We estimate the absorbed optical power through the flake by $I = I_0(1 - \exp(-\alpha d))$, where $\alpha$ is the absorption coefficient of WSe$_2$ and $d$ is the flake thickness. From DR spectroscopy measurements performed by Hsu et al.[4], we estimate the absorbance of single-layer WSe$_2$ for 640 nm to be 2%, corresponding to an absorption coefficient $\alpha = 0.4 \cdot 10^6$ cm$^{-1}$. In addition, we assume that all the incident laser power is converted into heat. Using the data in Figure 4b, we find

$$\eta_{\text{PV,max}} = 27.59\%, \qquad \eta_{\text{PTE,max}} = 0.0022\%.$$

Hence, the large absorbed optical power (6.35 µW compared to 1.57 nW) and small generated thermal voltage (106 mV compared to 732 mV) render power conversion by the photothermoelectric effect $\sim 1.5 \cdot 10^4$ times less efficient for power generation than the photovoltaic effect in the device.

**Supplementary references**